\input{aipcheck}

\documentclass[
    ,final            
  ]
  {aipproc}

\layoutstyle{6x9}

\newcommand{\lsim}{\mathrel{\mathop{\kern 0pt \rlap
  {\raise.2ex\hbox{$<$}}}
  \lower.9ex\hbox{\kern-.190em $\sim$}}}
\newcommand\gsim{\mathrel{\rlap{\lower4pt\hbox{\hskip1pt$\sim$}}
    \raise1pt\hbox{$>$}}}
\def\lsim{\mathrel{\raise.3ex\hbox{$<$\kern-.75em\lower1ex\hbox{$\sim$}}}}
\def\gsim{\mathrel{\raise.3ex\hbox{$>$\kern-.75em\lower1ex\hbox{$\sim$}}}}

\begin{document}

\title{UHECR Maps: mysteries and  surprises }

\classification{95.35.+d, 95.30Cq}
\keywords      {UHECR, composition, Nuclei, Local Group}

\author{D.Fargion}{
  address={Dip. di Fisica, Universit\`a di Roma ``Sapienza'', 00185 Rome, Italy},
  altaddress={INFN, sez. Roma 1, 00185 Rome, Italy}
}

\begin{abstract}
The rise of nucleon UHECR above GZK astronomy made by protons (AUGER November 2007) is puzzled by three main mysteries: an unexpected nearby Virgo UHECR suppression (or absence), a rich crowded clustering frozen vertically along Cen A, a composition suggesting nuclei (not much directional) and not nucleons. The UHECR map, initially consistent with  GZK volumes, to day seem to be not much correlated with expected Super Galactic Plane. Moreover slant depth data of UHECR from AUGER airshower shape do not favor the proton but points to  a nuclei, while HIRES, on the contrary favors mostly nucleons. We tried  (at least partially) to solve the contradictions assuming UHECR as light nuclei (mostly $He^4$) spread by planar galactic fields, randomly at vertical axis. The $He^4$ fragility and its mass and charge explains the Virgo absence (due to $He^4$ opacity above few Mpc) and the Cen A spread clustering  (a quarter of the whole sample). However more events and rare doublets and clustering elsewhere are waiting for an answer. Here we foresee hint of new UHECR component: galactic ones.  Moreover a careful updated  views of UHECR sky over different (Radio,IR,Optics, X,gamma, TeV) background (also Fermi gamma very last records) are also favoring forgotten  revolutionary Z-shower model. Both Z-Shower, proton GZK and Lightest nuclei UHECR models have dramatic influence on expected UHE neutrino Astronomy: to be soon revealed by UHE $\tau$ neutrino induced air-showers in different ways.

\end{abstract}

\maketitle


\section{INTRODUCTION}
The last century have seen the birth of a puzzling cosmic ray whose nature and origination has been and it is growing a never-ending chain of  puzzle.
The Cosmic Black Body Radiation had imposed since $1966$ a cut, GZK cut-off \cite{Greisen:1966jv}, of highest energy cosmic ray propagation, implying a very limited cosmic Volume (ten Mpc) for highest UHECR events. Because of the UHECR rigidity one had finally to expect to track easily UHECR directionality back toward the sources into a new Cosmic Rays Astronomy.
Indeed in last two decades, namely since 1991-1995 the rise of an \emph{apparent} UHECR at $3 {10^{20}}$ eV, by Fly's Eye,  has opened the wondering of its origination: no nearby (within GZK cut off) source have been correlated. Incidentally it should be noted that even after two decades  and after an increase of aperture observation  by nearly two order of magnitude (area-time by AGASA-HIRES-AUGER) no larger or equal event has been rediscovered. Making questionable by statistical argument the nature of that exceptional starting UHECR event. To face the uncorrelated UHECR at $3 {10^{20}}$ eV, the earliest evidences (by SuperKamiokande) that neutrino have a non zero mass
had opened  \cite{Fargion1997} the possibility of  an UHECR-Neutrino connection. An  UHE ZeV neutrino (born by a Seyfert galaxy MCG 8-11-11, which is very close to the arrival
direction of the Fly's Eye $3\cdot 10^{20}$ eV shower and it is located at a redshift of
z=0.0205, nearly 80 Mpc) could be the transparent  courier of that far AGN (beyond GZK radius) that may hit a wide local relic anti-neutrino dark halo, spread  at a few Mpc around our galaxy. Its Z boson (or WW channel) interaction \cite{Fargion1997} is source of a secondary nucleon observable at Earth as a UHECR.  The later search by AGASA seemed to confirm the Fly's Eye by events above $ {10^{20}}$eV  and the puzzling absence of nearby expected GZK anisotropy or correlation.  On $1999-2000$ we were all convinced on the UHECR GZK cut absence. However later records by a larger area Hires (2001-2005)  have been  reporting  evidences of GZK suppression. Sometimes with far BL Lac connections claim \cite{Gorbunov}. The same result seemed confirmed by last AUGER data in this a few years \cite{Auger-Nov07}. But in addition AUGER have shown an anisotropic clustering along the Super Galactic Plane, well consistent with GZK expectation \cite{Auger-Nov07}. The UHE neutrino scattering model \cite{Fargion1997},\cite{Weiler1997},\cite{Yoshida1998}) that had a wide audience in last decade, because of the smaller  recent neutrino mass limits (and the consequent  smaller relic neutrino number clustering and smaller interaction  probability) and because AUGER correlated map, became obsolete.
 In last two years the discrepancy with the composition, the absence of Virgo \cite{Gorbunov09} UHECR events and the clustering on Cen A, forced us and other to be suspicious  in UHECR-nucleon model and led us to consider lightest nuclei as the main UHECR events \cite{Fargion2008}. This solution explain part of the UHECR puzzle (Cen A cluster  and Virgo absence mainly), but more  uncorrelated UHECR events on different backgrounds need to be understood.
 Indeed here we review the most recent  UHECR maps \cite {Martello09} over different Universe \emph{colors} or (wave-length)  and we comment some surprising feature, noting some possible minor galactic component \cite{Fargion09b}. Moreover as shown in this paper with possible Z-Showering model solution. The consequences of the UHECR composition and source reflects into UHE (GZK \cite{Greisen:1966jv} or cosmo-genic) neutrinos. The proton UHECR provide EeV neutrinos (muons and electron) whose flavor oscillation lead to tau neutrinos to be soon detectable \cite{FarTau} \cite{Auger08} by upward tau air-showers;  the UHECR lightest nuclei model provide only lower energy, tens PeV, neutrinos detectable in a very peculiar way by AUGER fluorescence telescopes or in ARGO array by horizontal $\tau$ air-showers  , or by Icecube $km^3$ neutrino fluorescence telescopes \cite{FarTau},\cite{Fargion2009},\cite{Fargion09a}  \cite{Fargion09b} either by double bang\cite{Learned}, or long muon at few PeV energy. ZeV UHE neutrinos in Z-Shower model are possible source of horizontal Tau air-showers of maximal size and energy \cite{FarTau}.
\section{The Lorentz UHECR bending and spread}
Cosmic Rays are blurred by magnetic fields. Also UHECR suffer of a Lorentz force deviation. This smearing maybe source of UHECR features. Mostly along Cen A.
 There are two main spectroscopy of UHECR along galactic plane:
 A late nearby (almost local) bending by a nearest coherent galactic arm field, and a random one along the whole plane inside different arms.
The coherent Lorentz angle bending $\delta_{Coh} $ of a proton UHECR (above GZK \cite{Greisen:1966jv}) within a galactic magnetic field  in a final nearby coherent length  of $l_c = 1\cdot kpc$ is $ \delta_{Coh-p} \simeq{2.3^\circ}\cdot \frac{Z}{Z_{H}} \cdot (\frac{6\cdot10^{19}eV}{E_{CR}})(\frac{B}{3\cdot \mu G}){\frac{l_c}{kpc}}$.
The corresponding coherent  bending of an Helium UHECR at same energy, within a galactic magnetic field
  in a wider nearby coherent length  of $l_c = 2\cdot  kpc$ is
\begin{equation}
\delta_{Coh-He} \simeq
{9.2^\circ}\cdot \frac{Z}{Z_{He}} \cdot (\frac{6\cdot10^{19}eV}{E_{CR}})(\frac{B}{3\cdot \mu G}){\frac{l_c}{2 kpc}}
\end{equation}


This bending angle is compatible with observed multiplet along $Cen_A$ and also the possible clustering along Vela, at much nearer distances; indeed in latter case it is possible for a larger magnetic field along its direction (20 $\mu G$) and-or for a rare iron composition $\delta_{Coh-Fe-Vela} \simeq
{17.4^\circ}\cdot \frac{Z}{Z_{Fe}} \cdot (\frac{6\cdot10^{19}eV}{E_{CR}})(\frac{B}{3\cdot \mu G}){\frac{l_c}{290 pc}}$. Such iron UHECR are mostly bounded inside a Galaxy, as well as in Virgo, explaining partially  its extragalactic absence. In lightest nuclei model the heavier of lightest nuclei that may be bounded from Virgo, Be, is bent by $
\delta_{Coh-Be} \simeq
{18.4^\circ}\cdot \frac{Z}{Z_{Be}} \cdot (\frac{6\cdot10^{19}eV}{E_{CR}})(\frac{B}{3\cdot \mu G}){\frac{l_c}{2 kpc}}
$.  The incoherent random angle bending, $\delta_{rm} $, while crossing along the whole Galactic disk $ L\simeq{20 kpc}$  in different spiral arms  and within a characteristic  coherent length  $ l_c \simeq{2 kpc}$ for He nuclei is
$
\delta_{rm-He} \simeq
{16^\circ}\cdot \frac{Z}{Z_{He^2}} \cdot (\frac{6\cdot10^{19}eV}{E_{CR}})(\frac{B}{3\cdot \mu G})\sqrt{\frac{L}{20 kpc}}
\sqrt{\frac{l_c}{2 kpc}}
$
The heavier  (but still lightest nuclei) bounded from Virgo are Li and Be:
$\delta_{rm-Li} \simeq
{24^\circ}\cdot \frac{Z}{Z_{Li^3}} \cdot (\frac{6\cdot10^{19}eV}{E_{CR}})(\frac{B}{3\cdot \mu G})\sqrt{\frac{L}{20 kpc}}
\sqrt{\frac{l_c}{2 kpc}}
$, $
\delta_{rm-Be} \simeq
{32^\circ}\cdot \frac{Z}{Z_{Be^4}} \cdot (\frac{6\cdot10^{19}eV}{E_{CR}})(\frac{B}{3\cdot \mu G})\sqrt{\frac{L}{20 kpc}}
\sqrt{\frac{l_c}{2 kpc}}
$.  It should be noted that the present anisotropy above GZK \cite{Greisen:1966jv} energy $5.5 \cdot 10^{19} eV$ might leave a tail of signals: indeed the photo disruption of He into deuterium, Tritium, $He^3$ and protons (and unstable neutrons), might rise as clustered events at half or a fourth of the energy.  It is important to look for correlated tails of events, possibly in  strings at low $\simeq 1.5-3 \cdot 10^{19} eV$ along the $Cen_A$ train of events. \emph{It should be noticed that Deuterium fragments are half energy and mass of Helium: Therefore D and He spot are bent at same way and overlap into circle clusters.}In conclusion He like UHECR  maybe bent by a characteristic as large as  $\delta_{rm-He}  \simeq 16^\circ$. Well within the observed CenA UHECR clustering spread.


\section{UHECR Maps: a brief tour in multi-wave sky}
\begin{figure}[!ht]
\includegraphics[width=7cm]{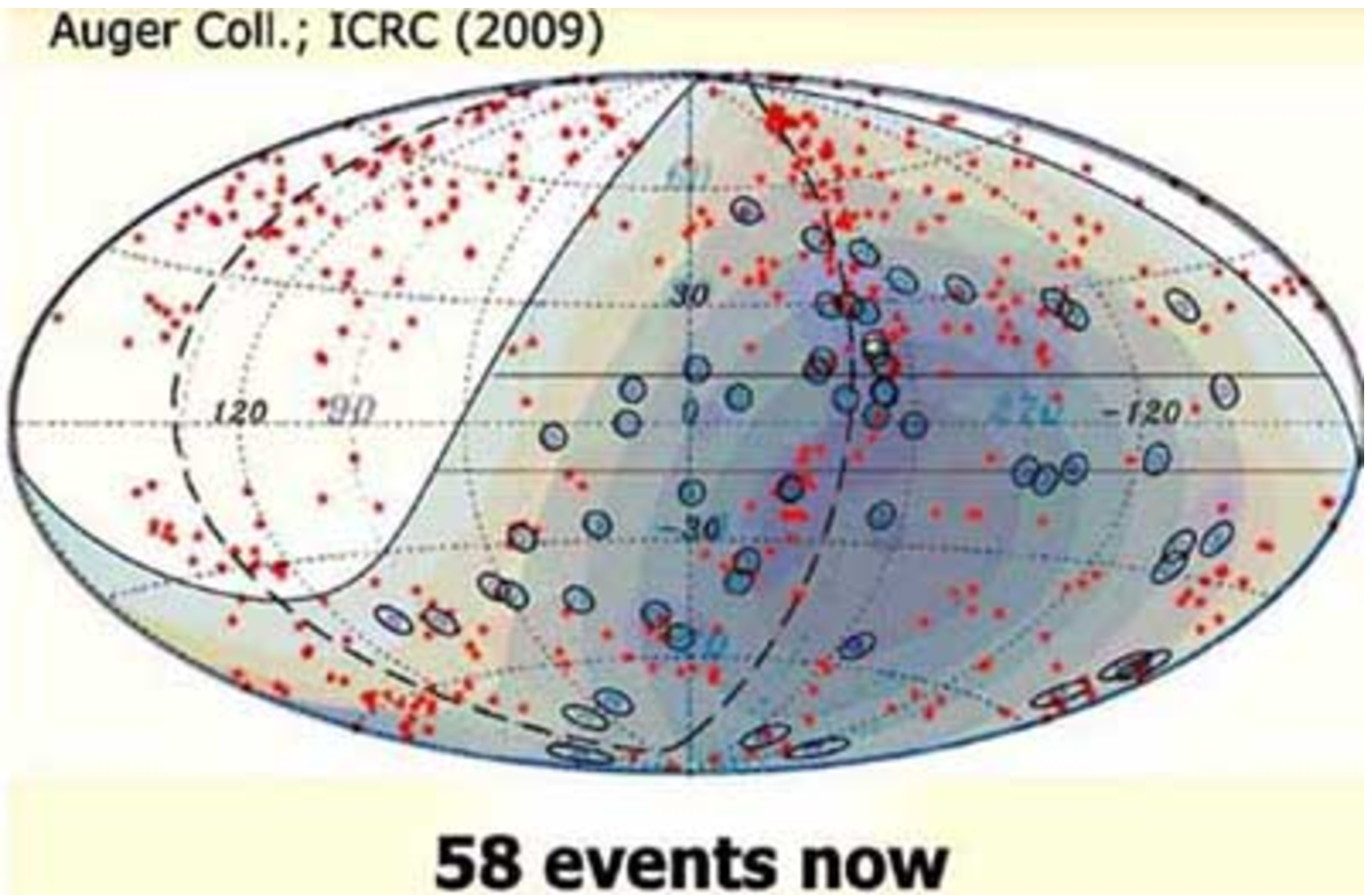}
\includegraphics[width=7cm]{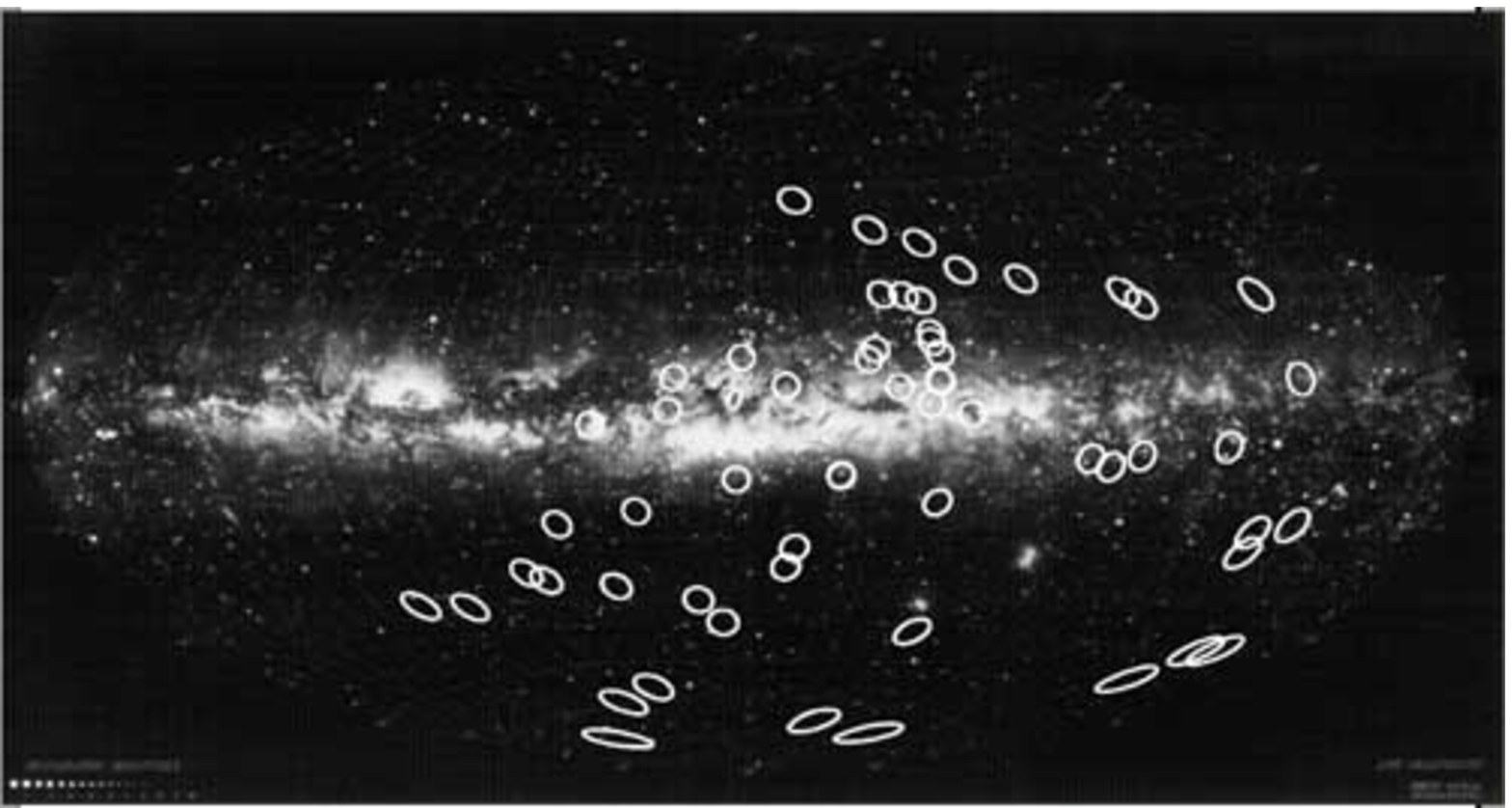}
\vspace{0.3cm}
\caption{Left: An overlap of ICRC09 and Scineghe09 presentation maps. The two UHECR event description are self consistent and have been used as a mask in present and early study
see \protect\cite{Fargion09a},\protect\cite{Martello09}; the probability to \emph{not} observe any UHECR in Virgo area within (one  $3.5^{o}$,two $7^{o}$, or three $10.5 ^{o}$ angle radius) far from each AGN (small red star, within the same less probable AUGER sky area), is respectively $37.15\%, 1.9\%, 0.12\%$; therefore it is clear from these unlike statistics the puzzling paucity of (expected) Virgo sources.  Right: a visible sky in galactic coordinate within recent 58 UHECR events. The ring radius is nearly $3.5^o$. No apparent correlation between lights and UHECR records maybe found in this map.
see \protect\cite{Fargion09a},\protect\cite{Martello09}.}
\label{fig1}
\end{figure}

We offer in next figures a map view of the last AUGER UHECR events over the different sky wave-band astronomy. At each step the caption explains and offer the arguments for mysteries and surprises. We begin on figure \ref{fig1} to show our mask calibration based on two AUGER different recent presentation (HEP 2009, Scineghe 2009 \cite{Martello09} ) and the same event are shown on optical sky in common galactic coordinate. In the next figure \ref{fig2} we discuss the two micron view and we make a few consideration on a probable galactic (secondary) UHECR component (to be confirmed in a near future). Therefore we overlay these UHECR ring events onto a recent Local Universe Map showing its viability and success to correlate with most events ( figure \ref{fig3}). The Lightest nuclei model naturally explain the need of such a nearby Universe for UHECR.
However the peculiar spread of the events around Cen A makes us believe that vertical spread is not only related to galaxies distribution but also on Lorentz bending. The next  ( figure \ref{fig4}) describe the UHECR over the $408$ Mhz cosmic background, showing the peak role of the Cen A emission and Vela emission with UHECR clustering. The following  five maps  ( figure \ref{fig5}, figure \ref{fig7},figure \ref{fig9}), are dealing with extragalactic cosmic volume ($0-40; 40-80; 80-120; 120-160; 160-200$ Mpc) within GZK one for a proton UHECR. The arguments showing mysteries, contradictions and puzzle are commented. The overall picture is not much in agreement with expected UHECR made by nucleon. In the ( figure \ref{fig10}) we show the most recent composition derived by slant depth of the UHECR air-showers both for Hires and AUGER, as well as for HIRES and MIA earlier and late records, all over composition model for UHECR. The possibility for the lightest nuclei (mostly Helium) to council   both experiments results is manifest. In next ( figure \ref{fig11}) we continue to show the sky at highest photon energies: a X sky and the most famous label sources are overlay with UHECR; no clear evidence is derived, but some comment and hints are suggested. The oldest  EGRET maps with labels and some  correlations with UHECR are shown ( figure \ref{fig12}). The next figure ( figure \ref{fig13}) consider the surviving fraction of proton with distances as well as the interaction distance for nucleon and lightest nuclei with remarks on Lightest nuclei solution. The final  ( figure \ref{fig14}) overlap the very hard Fermi and TeV sky with UHECR events: a clear Cen A role is blowing, but  also few far AGN or BL Lac are correlating, most well above GZK cut-off. In particular in last figure it is shown the last (second December 2009) dramatic bright gamma shining of the AGN  $3C454.3$ whose last flare at cosmic  edge ($ redshift  z=0.859 $ well above half size Universe) and its eventual correlation might force us to a Z-Shower solution .

\begin{figure}[!ht]
\includegraphics[width=9cm]{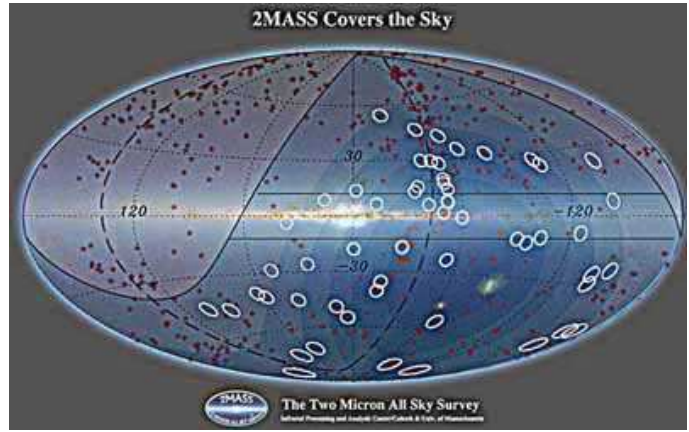}
\caption{The two micron sky, a near infrared view that do not correlate with the observed AUGER UHECR events. Note however the quadruplet
events around the galactic center hinting a galactic component. Also consider a triplet along the galactic plane possibly related to Vela, the nearest and brightest gamma and radio source.
see also \protect\cite{Fargion09a}. and in particular see\protect\cite{Fargion09b}}
\label{fig2}
\end{figure}

\begin{figure}[!ht]
\includegraphics[width=9cm]{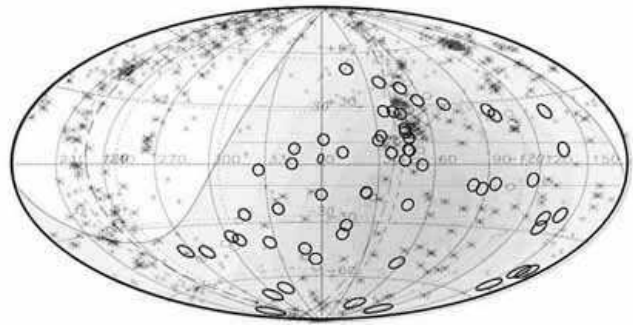}
\caption{The very near Universe galaxies (marked by black stars), within few Mpc , maximally correlated with earliest $27$ UHECR (see \protect\cite{Local09}  and the most recent events by AUGER);(to not be confused with the background GZK AGN  assumed by AUGER as primary sources, marked by smaller red stars). It  is manifest the strong correlation \emph{vertically} along the Cen A region and a partial correlation elsewhere. The absence of UHECR on connection with the  cluster on the top right side is possibly related to the larger distance and bias detection. Lightest nuclei explain naturally such a successful  narrow preference of UHECR view. However the inner clustering centered on Cen A favor in my opinion, an unique source and a galactic field bending of UHECR.
see \protect\cite{Fargion09a},\protect\cite{Fargion09b}}
\label{fig3}
\end{figure}

\begin{figure}[!ht]
\includegraphics[width=9cm]{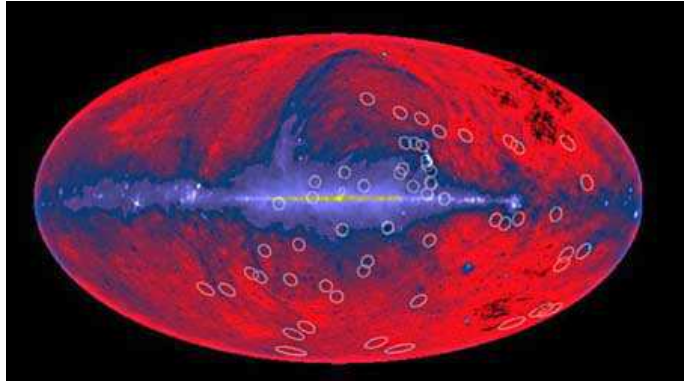}
\caption{The $408$ Mhz sky and the AUGER UHECR events. The peak activity of Cen A in radio (and partially in gamma) and its UHECR clustering hint a similar role for nearest Vela PSR, whose radio and gamma activity is brightest and extreme.
see \protect\cite{Fargion09a},\protect\cite{Fargion09b}}
\label{fig4}
\end{figure}

\begin{figure}[!ht]
\includegraphics[width=9cm]{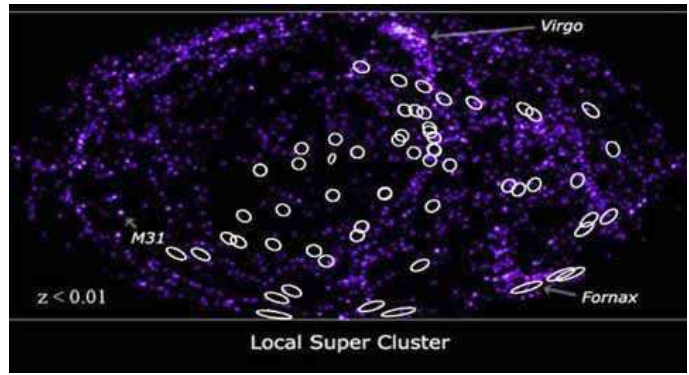}
\caption{ The nearby ($0-40$) Mpc Universe and the UHECR AUGER events. It is well apparent the bright cluster of Virgo remarked by the absence of any UHECR. The UHECR GZK proton cut-suppression reduces the Virgo UHECR flux by a negligible  factor $\simeq 20\%$, leading to an expected  factor $\simeq 80\%$ bright signal yet (as long as it is public) not observed. It maybe noted that AUGER Virgo sky is not the best to reveal by AUGER see \protect\cite{Auger-Nov07}. However  a simple statistical Poisson test (based on a sample of $58$ UHECR events counts  at same equiprobable area) imply a small probability  $\simeq 1.9\%$ to not find any signal in Virgo area or even less $1.2 10^{-1}\%$ allowing $13.5^{o}$ angular dispersion around each AGN (potential sources); this count is based on the AGN presence in the Virgo direction over other AGN in equiprobable AUGER areas. From this paradox our He-UHECR model explains at best the Virgo absence by severe lightest nuclei opacity, while coexisting with a nearer Cen A UHECR clustering . The Fornax cluster presence might be related to nearer galaxy signals.
see \protect\cite{Local09}.\protect\cite{Fargion09a},\protect\cite{Fargion09b}}
\label{fig5}
\end{figure}

\begin{figure}[!ht]
\includegraphics[width=7cm]{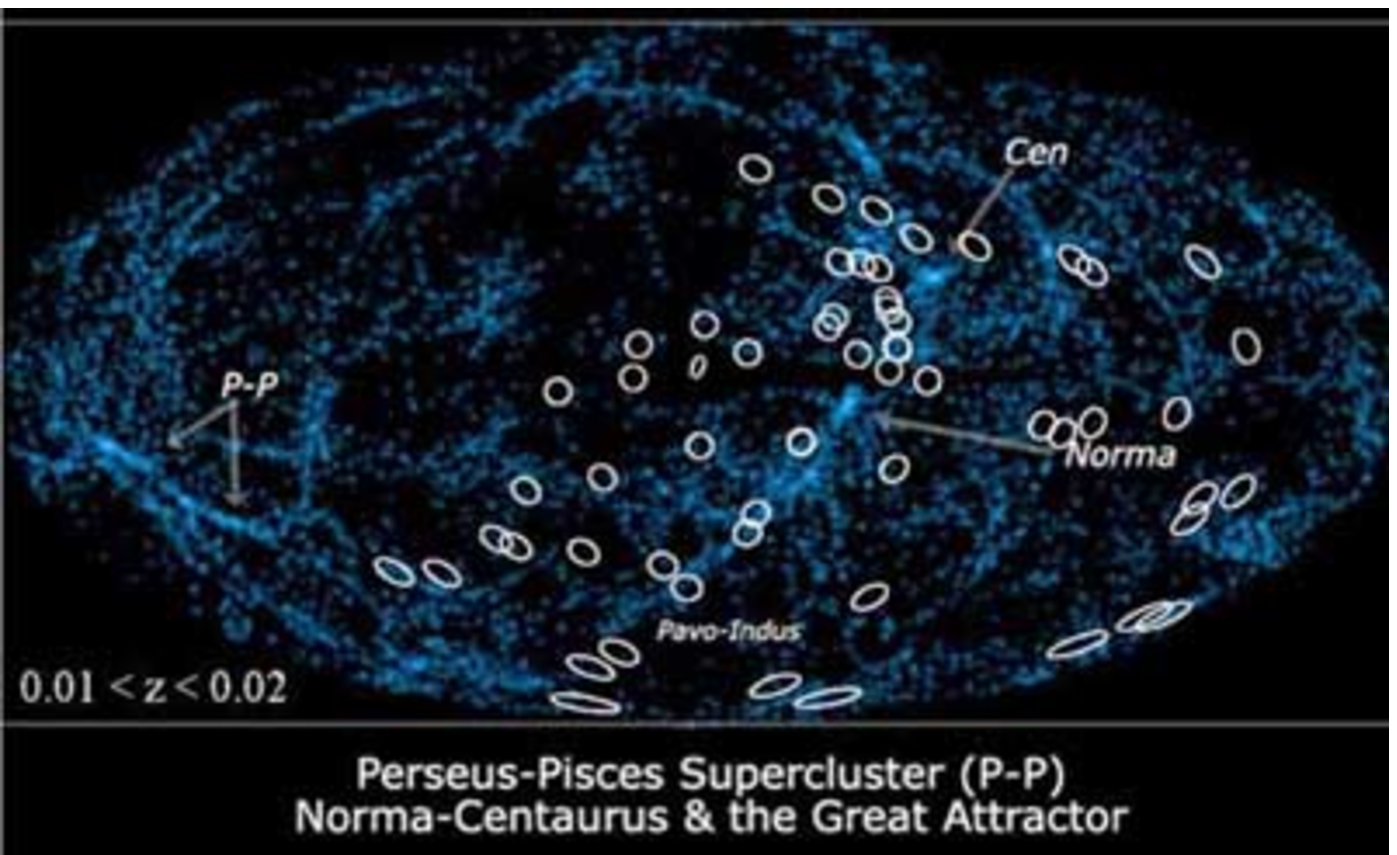}
\includegraphics[width=7cm]{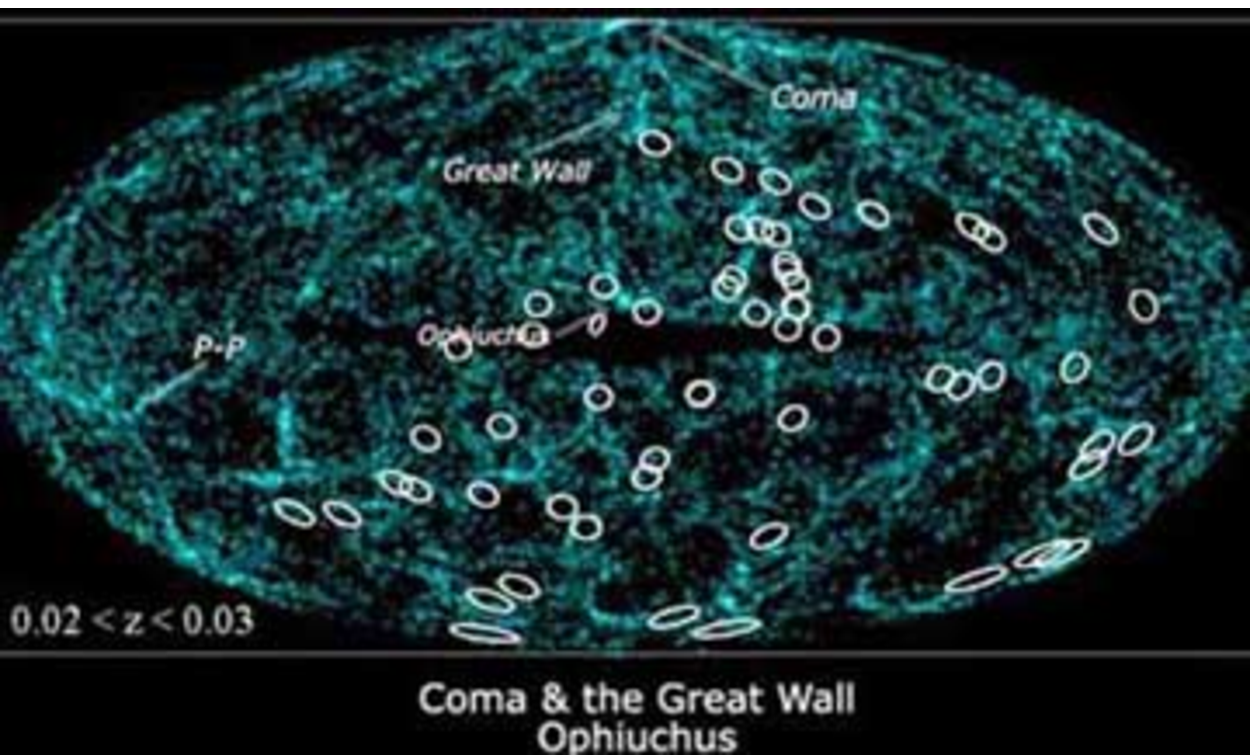}
\vspace{-0.3cm}
\caption{Left: This map consider a wider GZK volumes at better correlation with UHECR. The distances ranges between ($40-80$) Mpc, and Virgo is no longer ruling, Therefore there is here an (apparent) promising agreement between records and nucleons UHECR expectations. Nevertheless even here there are two manifest contradictions: the displacement of Cen Cluster from the main Cen A crowded AUGER events, the absence of Norma enhancement just where the AUGER detection is maxima. Moreover the GZK opacity for nucleon  here absorb more the UHECR arrival reducing to $\simeq 67\%$-$\simeq 45\%$ their transmission.
 Right: In this volumes wider GZK volumes,  between ($80-120$) Mpc, the correlation between UHECR,  Ophiuchus cluster and Great Wall, as above,  seem promising for proton UHECR model. The spread of IR sources might correlate somehow more UHECR events. However the strong Centaurus correlation is lost and the GZK opacity is relevant reducing to $\simeq 45\%$-$\simeq 25\%$ the UHECR transmission.}
\label{fig7}
\end{figure}

\begin{figure}[!ht]
\includegraphics[width=7cm]{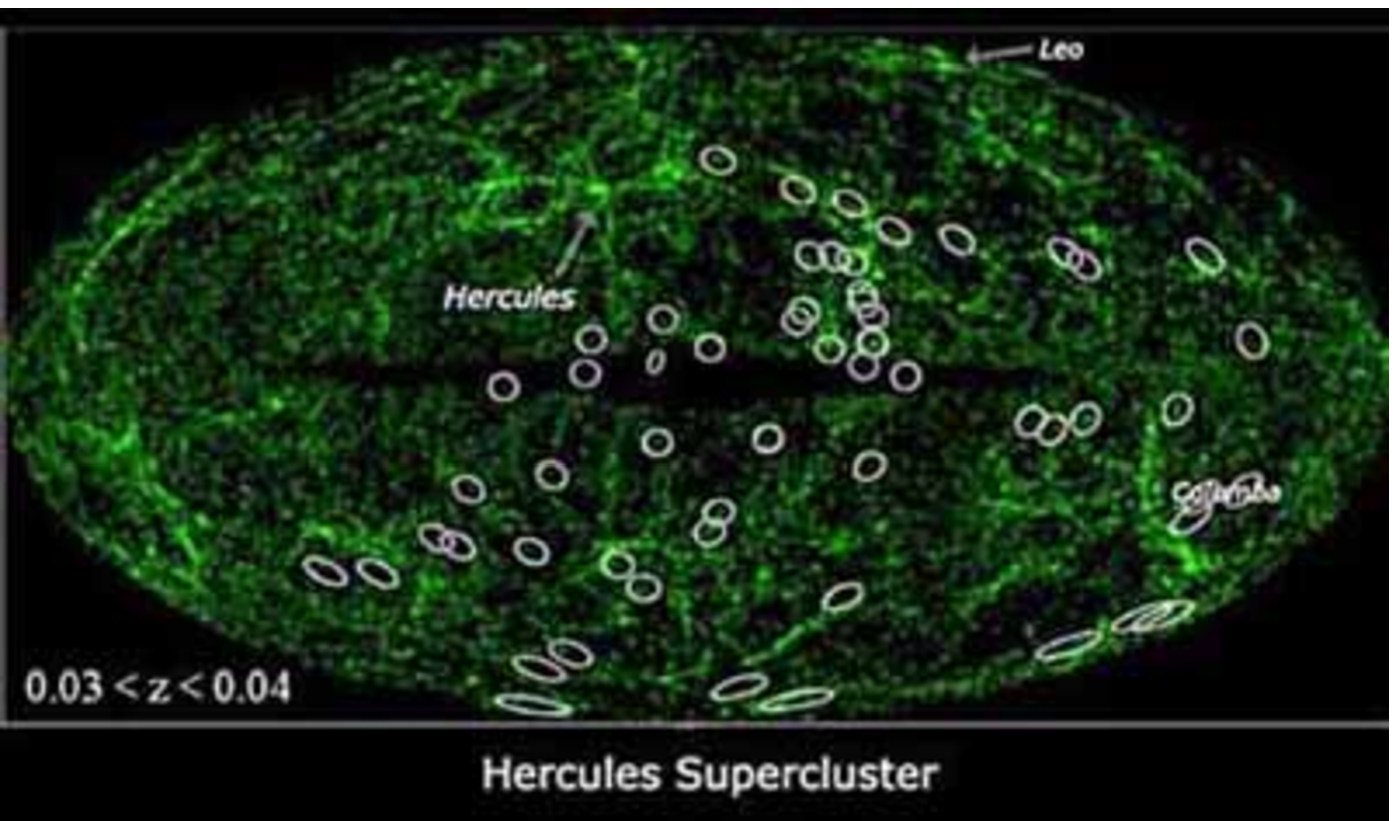}
\includegraphics[width=7cm]{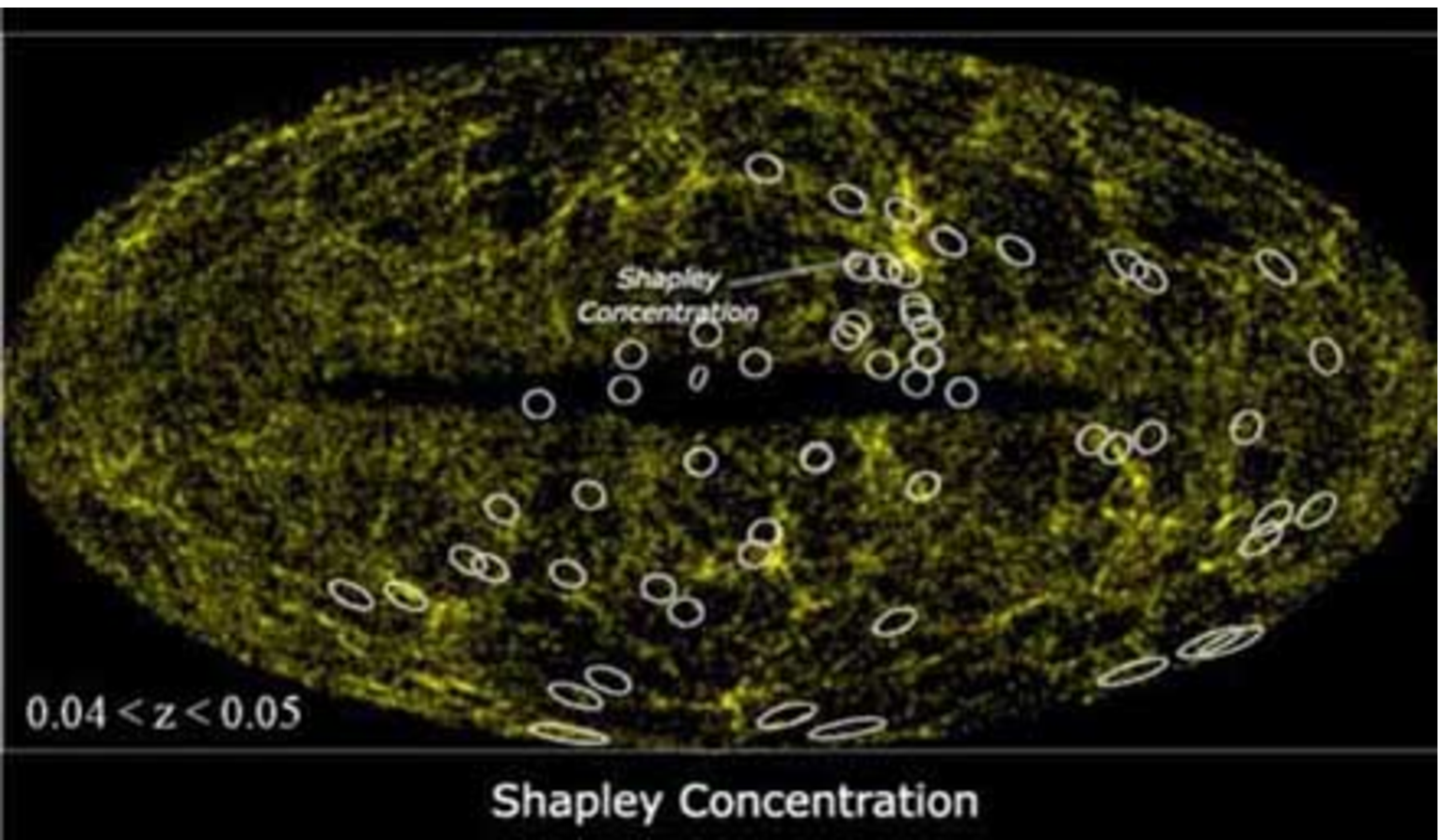}
\vspace{-0.3cm}
\caption{Left: In this GZK extreme edge there is some approximate correlation with the UHECR map: Columba supercluster may explain two triplets at extreme regions, while Hercules absence maybe related to AUGER thresholds. However at those distances  between ($120-160$) Mpc, the GZK opacity is severe:$\simeq25\%$-$\simeq 10\%$ for the UHECR transmission. Right: In this very GZK extreme edge, between ($160-200$) Mpc distances, the map seem to fit at best with the UHECR records. The Shapley concentration and other minor clusters  seem to overlap in a resonant way the observed AUGER UHECR events. However in a proton model the GZK opacity is dramatic: only $\simeq10\%$-$\simeq 5\%$ of the events might be transmitted to us. Such a reduction must be added to the distance spread area, making, if not unrealistic, at least difficult also such proton UHECR understanding. In particular also for the absence of Virgo sources.
}
\label{fig9}
\end{figure}

\begin{figure}[!ht]
\includegraphics[width=8cm]{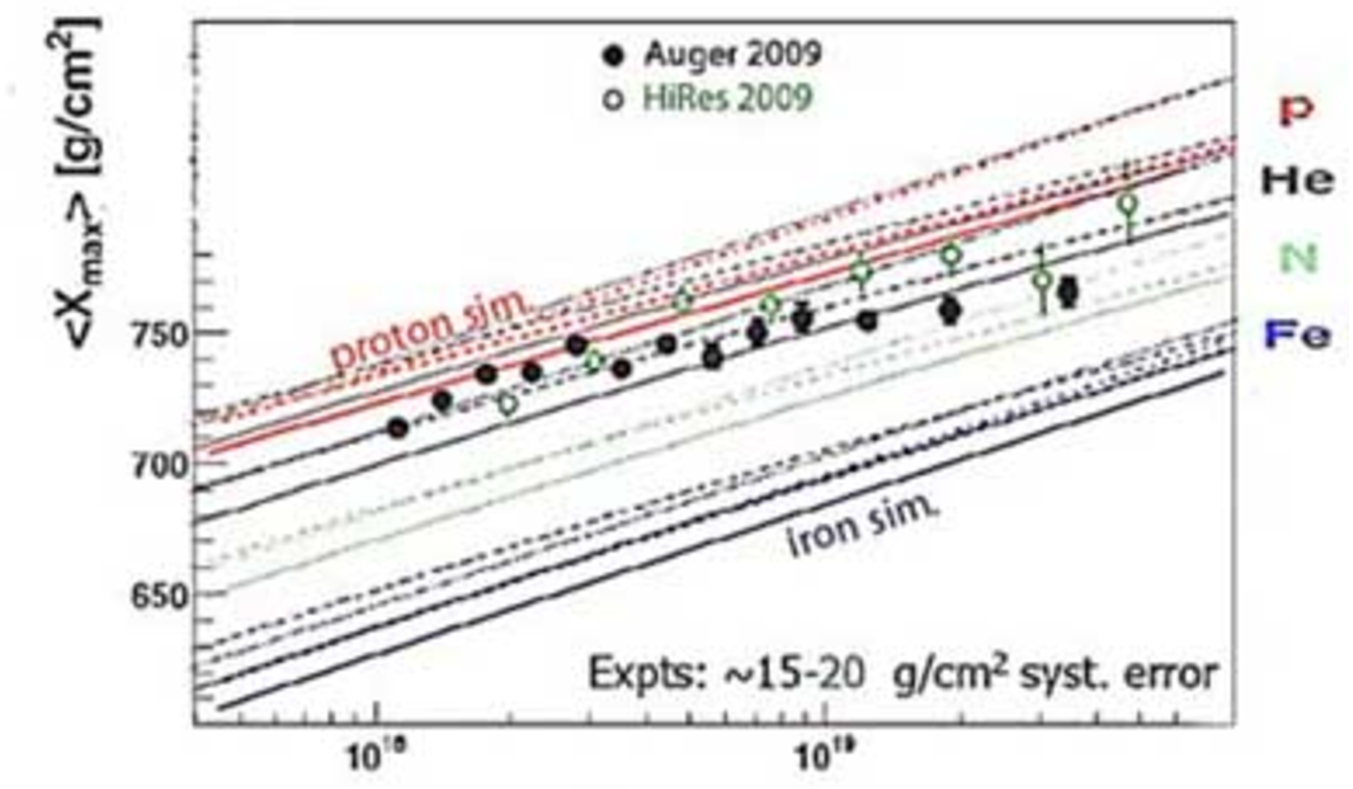}
\includegraphics[width=8cm]{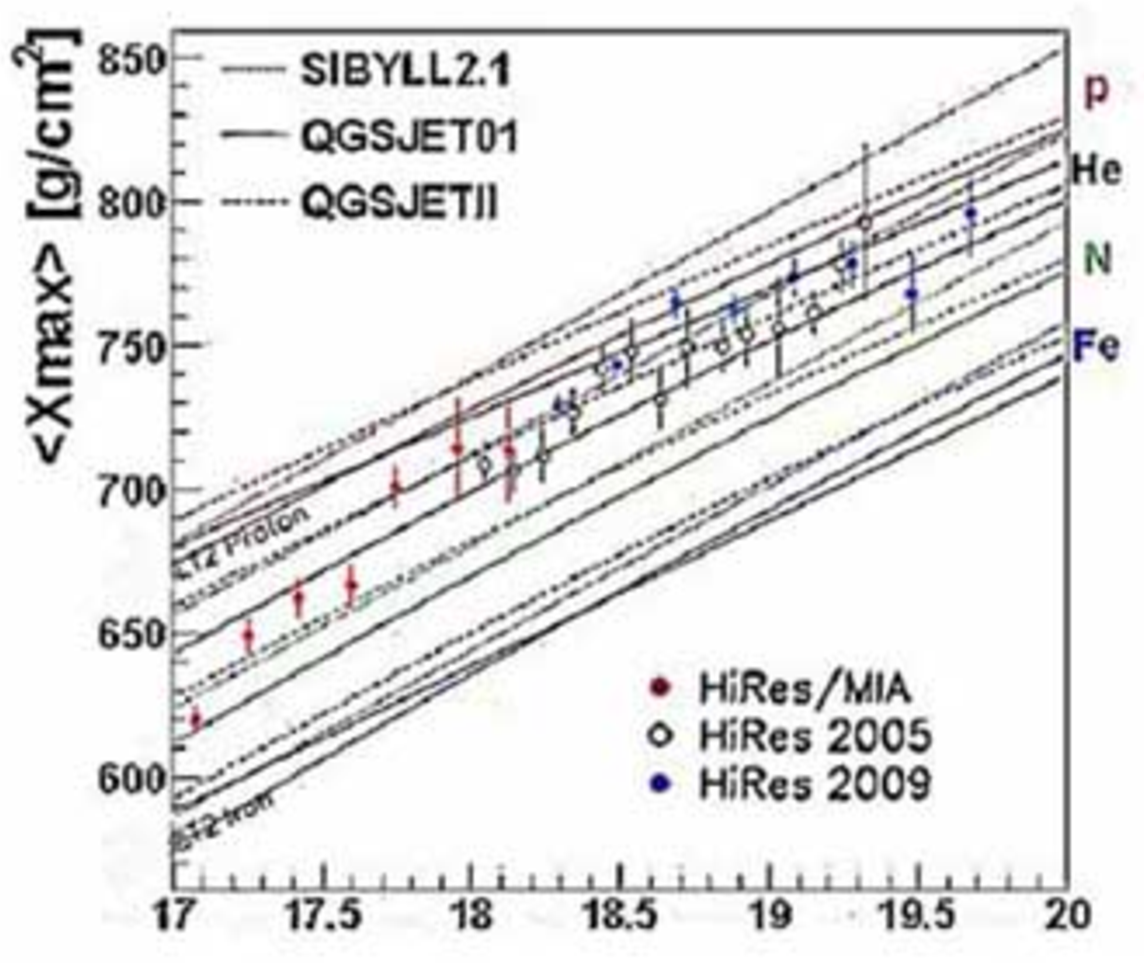}
\vspace{-0.3cm}
\caption{Left: The most recent  UHECR composition map derived by AUGER and HIRES with expected model curves. The He possibility is a better solution also respect an iron-proton mix, because of the narrow error band in AUGER records. Right: The most recent  UHECR composition map derived by  HIRES and Hires-MIA in last decade with expected model curves. The He possibility is a viable solution.
}
\label{fig10}
\end{figure}

\begin{figure}[!ht]
\includegraphics[width=10cm]{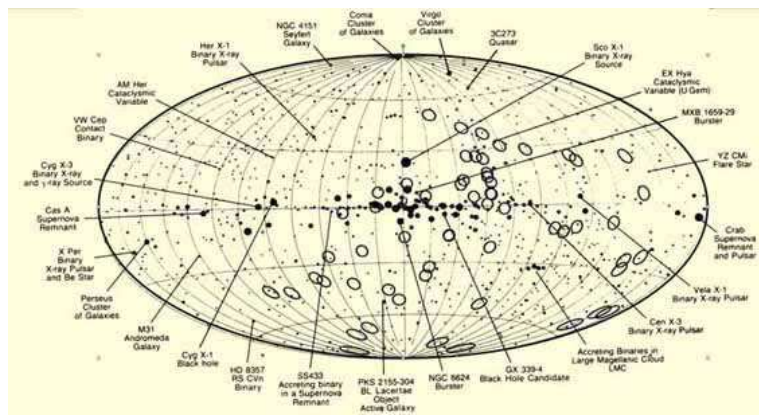}
\vspace{-0.3cm}
\caption{ The earliest well known X-Ray sources over AUGER $58$ UHECR events. The absence of any strong galactic signature is somehow tempered by the presence of some correlation between X-sources and UHECR around North galactic center. Peculiar sources as spinning, precessing Jet SS433, might correlate. Such a microquasar source at brightest activity maybe ejecting UHECR. Some extragalactic AGN, to discuss later, also correlate.
}
\label{fig11}
\end{figure}
\begin{figure}[!ht]
\includegraphics[width=9cm]{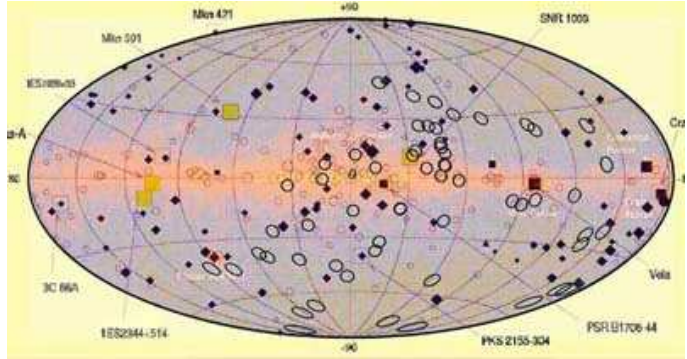}
\vspace{-0.3cm}
\caption{ The old gamma EGRET map with remarkable sources name. It is worth to notice: a few correlated UHECR events rings along the galactic center, the SN1006 source, but also some surprising potential connection of a doublet with far extragalactic AGN as $PKS 2155-304$ and an unexpected far $3C454.3$ whose distance is almost half the Universe size from us. Such a rare connection, well above GZK distance by more than ten times recall the revolutionary (but un-fashion) courier role of  UHE neutrino at ten ZeV energy: they may hit onto relic neutrinos on the way (within GZK radius) and lead to Z-shower whose nucleons may be the  observed signal, see \protect\cite{Fargion1997},\cite{Weiler1997};\cite{Yoshida1998}.
}
\label{fig12}
\end{figure}

\begin{figure}[!ht]
\includegraphics[width=6cm]{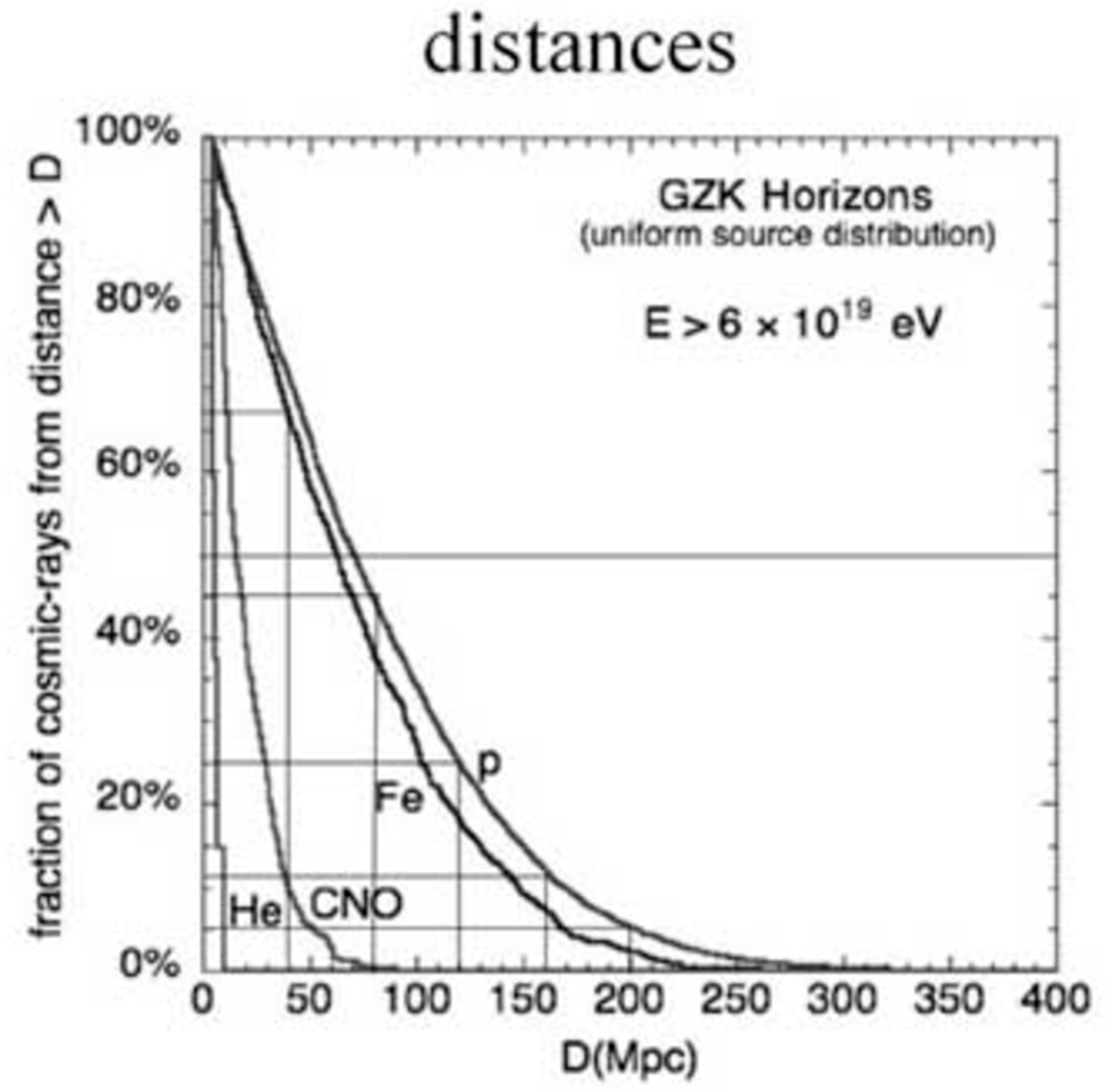}
\includegraphics[width=7cm]{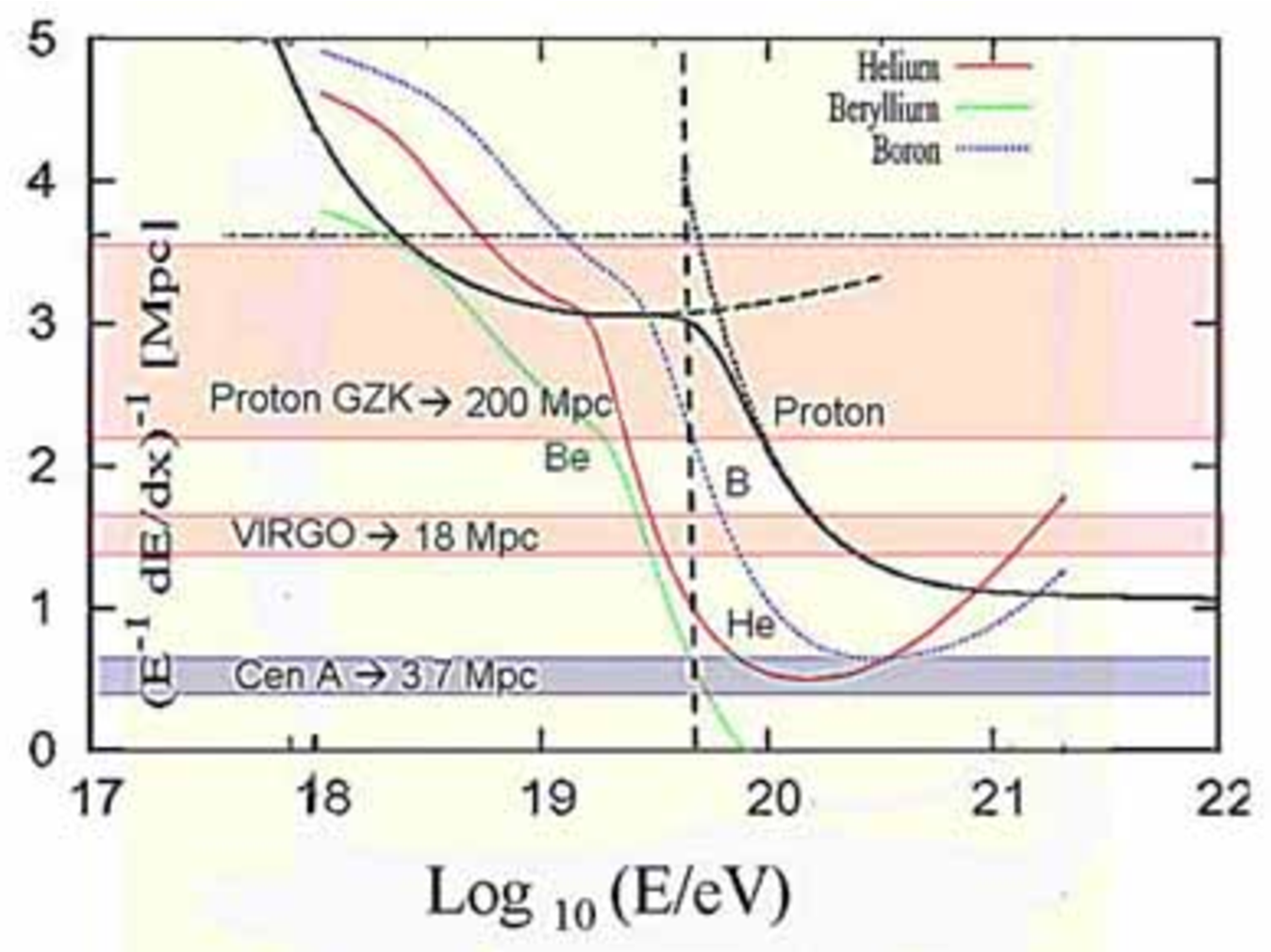}
\vspace{-0.3cm}
\caption{ Left: the suppression fraction of UHECR proton with distance \cite{Auger-Nov09} : this distance dependence explain why UHECR $He^{4}$ are unable to reveal Virgo and why different Infrared maps above are more or less opaque to GZK. The extreme far AGN at far red-shift ($z\geq 0.4$)  are exponentially suppressed by GZK cut-off. This explain the merit of the UHECR originated by UHE neutrino scattering at Z-resonance \protect\cite{Fargion1997},\cite{Weiler1997};\cite{Yoshida1998}.
On the right side the interaction length for proton, Helium and lightest nuclei, showing the main extreme distances. This curve explain why , for Helium like  nuclei, Cen A maybe well observed while Virgo is already isolated from us. }
\label{fig13}
\end{figure}
\begin{figure}[!ht]
\includegraphics[width=8.1cm]{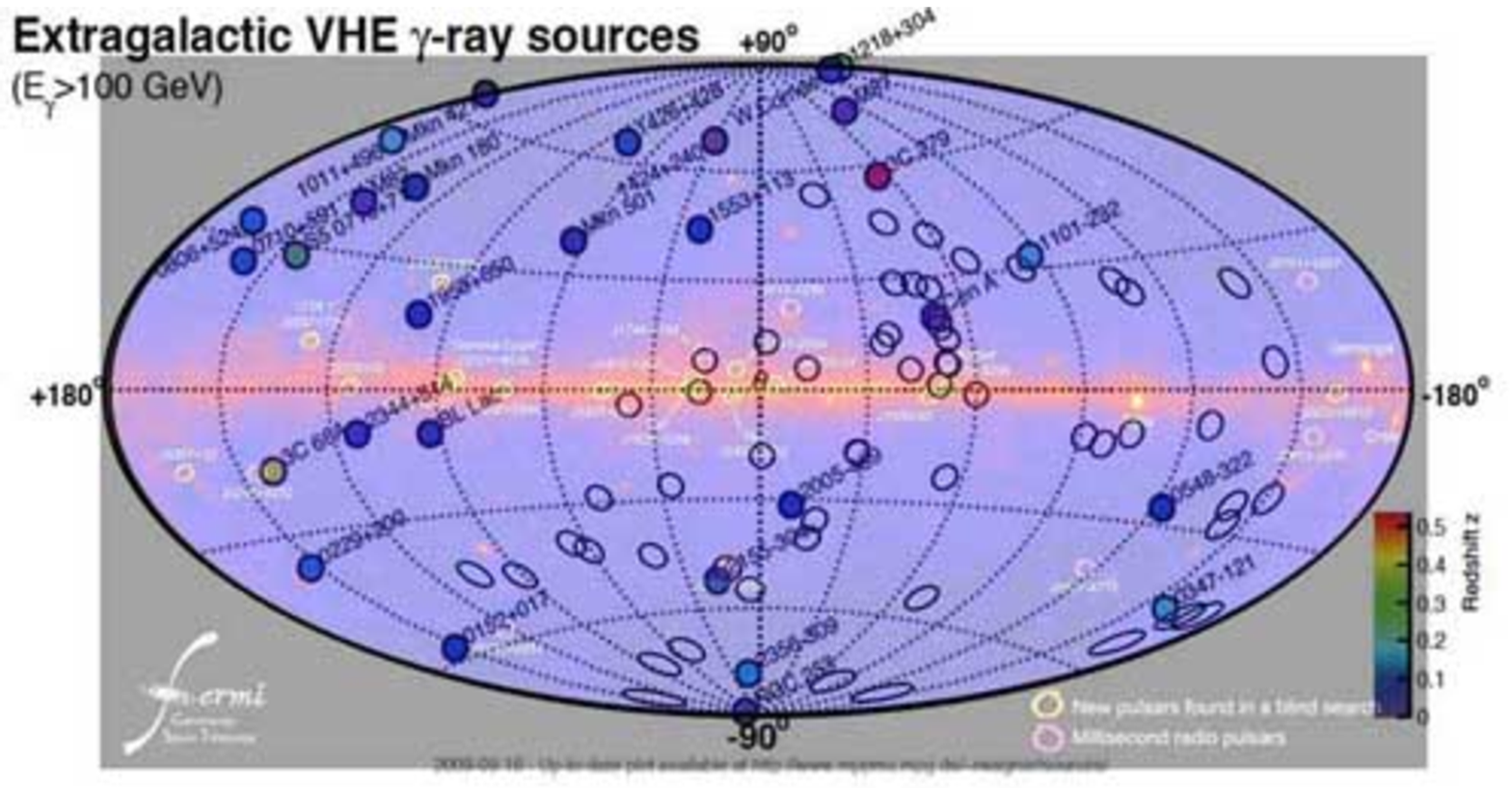}
\includegraphics[width=6.8cm]{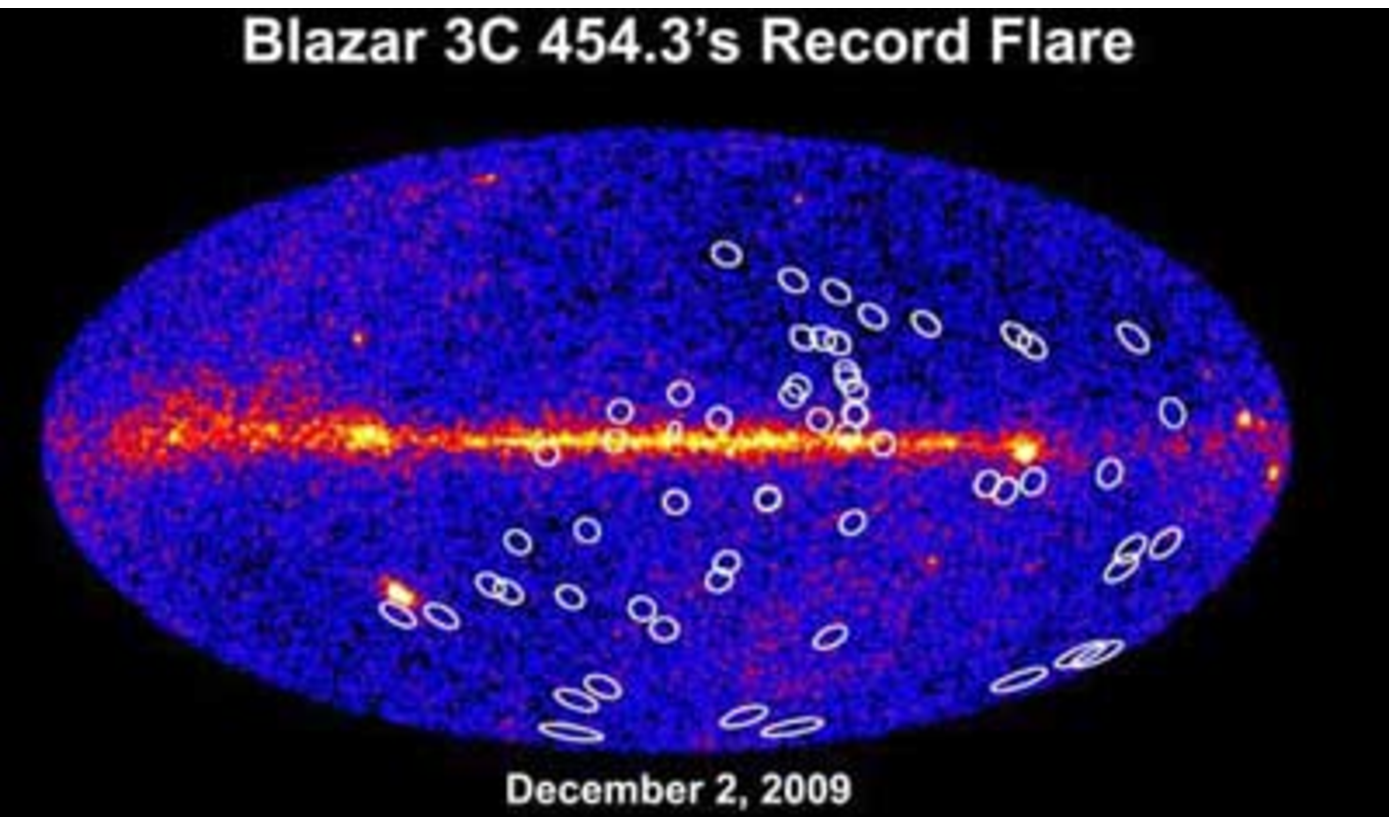}
\vspace{-0.3cm}
\caption{Left: The Hardest Gamma sources versus AUGER maps. Under the image is shown the Fermi gamma sky in low background. In this sky dot map, (whose blue-red colors mark the nearby-far red-shift), one observe the remarkable Cen A correlation with the main UHECR string events. However one is tempted to correlate also the AGN $1101-282$, AGN $0347-121$,AGN $2005-489$,AGN $2155-304$. These sources are (by redshift colors) well above the GZK cut off. The only viable possibility is an extreme UHE neutrino at ZeV energies scattering on relic ones within a GZK volumes \protect\cite{Fargion1997},\cite{Weiler1997};\cite{Yoshida1998}. The very relevant UHECR connection is with blazar $3C454.3$ whose exceptional flare has been discovered just last month. Its distance is half the way the Universe (above two Gpc) requiring (if the correlation is true) a an extreme UHE neutrino at ZeV energies scattering on relic ones. See next figure. Right:The very recent Flare from far AGN blazar $3C454.3$, at half the Universe distance in Fermi sky and with UHECR. Its relevance is not just related to the huge output of the source and to the doublet AUGER event connected by this map: but also to additional signals to be discussed in forthcoming article. The Z-resonant (or Z-Burst) model explains this otherwise mysterious connection. The UHE neutrino primary energy need to be nearly $10-30$ ZeV and the relic neutrino mass might range in the $0.4-0.133$ eV. The whole conversion efficiency might range from a minimal $10^{-4}$ \cite{Fargion1997} for no relic neutrino clustering up to $4 \cdot 10^{-3}$ a (forty) times density contrast in Local Group halo. Even within the minimal conversion efficiency, observed  gamma flaring $3C454.3$ blazar  is consistent with the extreme UHECR flux assuming a primary Fermi flat spectra (of the blazar) extending up to ZeV energy. See \cite{Fargion1997}.
}
\label{fig14}
\end{figure}
\section{Conclusions}
 The history of Cosmic Rays and last UHECR discoveries (and disclaims) are exciting and surprising. The list of models that rose and fall  just last decade is confusing but also promising of new horizons and revolutions. We all hope with courageous experiment fighters on the ground, as Fly's Eye, AGASA, Hires and AUGER ones, are no disappointed and that a new UHECR  astronomy is born, possibly as expected within a GZK Universe \cite{Auger-Nov09}. But Nature is sometime hides its final picture inside inner boxes. The very surprising correlation with Cen A, the absence of Virgo, the hint of correlation with  Vela and galactic center might be solved by a lightest nuclei, mainly He, as a  courier, leading to a very narrow (few Mpc) sky for UHECR. However the very  exceptional  blazar $3C454.3$ flare on $2nd December$ $2009$, a month ago, and the few AGN connection of UHECR far from a GZK volume may force us, surprisingly, to reconsider an exceptional model: Z-Shower one. Possibly connecting lowest neutrino   particle ($\simeq 0.15$ eV) mass with highest UHE ($\simeq 30$ZeV)neutrino energies \cite{Fargion1997}. Even for the minimal UHE $\nu$-Z-UHECR conversion as low as $10^{-4}$ (see table$1$,last reference in\cite{Fargion1997}), \emph{even for a not clustered relic neutrino halo as diluted as cosmic ones}  the present gamma $3C454.3$  output (above $3\cdot 10^{48}$ $erg s^{-1}$) is  comparable with UHECR (two events in $4$ years in AUGER), assuming  a flat Fermi spectra (for neutrinos) extended up to UHECR ZeVs edges. These results are somehow surprising and revolutionary. We might be warned for unexpected very local and a very wide  Universe sources sending UHECR traces in different ways. Testing for the first time the most evanescent \emph{hot} neutrino relic background. The consequences may be soon detectable in different way by Tau air-showers in AUGER,TA, Heat and also in unexpected horizontal shower in deep valley by ARGO \cite{FarTau},\cite{Auger08}. Or  in widest atmosphere layer on Earth, Jove and Saturn, see 2007 papers in\cite{Fargion1997}. Additional surprises, to be discussed in detail elsewhere, are  waiting beyond the corner. A soon answer maybe already written into present  clustering (as Deuterium fragments) at half UHECR edge energy around or along main UHECR group seed. Better understanding will rise by new data. The next release of UHECR update events and maps may solve the puzzles.

\end{document}